\documentclass[prb,twocolumn,showpacs,superscriptaddress,preprintnumbers,amsmath,amssymb,showkeys]{revtex4}

\usepackage{epsfig}
\usepackage{graphicx}
\usepackage{color}
\usepackage{amsmath}

\begin{document}

\newcommand{\SiOtwo}{SiO$_2$}
\newcommand{\SiOx}{SiO$_x$}
\newcommand{\Ntwo}{N$_2$}
\newcommand{\degree}{^{\circ}}
\newcommand{\Rglass}{R_{\rm glass}}

\title{Nanophotonic hybridization of narrow atomic cesium resonances and photonic stop gaps of opaline nanostructures}

\author{Philip J. Harding}
\affiliation{Complex Photonic Systems (COPS), MESA$^+$ Institute for Nanotechnology, University of Twente, P.O. Box 217, 7500 AE Enschede, The Netherlands}
%\email{p.j.harding@alumnus.utwente.nl}

\author{Pepijn W.H. Pinkse}
 \email{P.W.H.Pinkse@utwente.nl}
\affiliation{Complex Photonic Systems (COPS), MESA$^+$ Institute for Nanotechnology, University of Twente, P.O. Box 217, 7500 AE Enschede, The Netherlands}
%\affiliation{Applied Nanophotonics (ANP), MESA$^+$ Institute for Nanotechnology, University of Twente, P.O. Box 217, 7500 AE Enschede, The Netherlands}

\author{Allard P. Mosk}
\affiliation{Complex Photonic Systems (COPS), MESA$^+$ Institute for Nanotechnology, University of Twente, P.O. Box 217, 7500 AE Enschede, The Netherlands}
%\email{A.P.Mosk@utwente.nl}

\author{Willem L. Vos}
\email{W.L.Vos@utwente.nl}
\affiliation{Complex Photonic Systems (COPS), MESA$^+$ Institute for Nanotechnology, University of Twente, P.O. Box 217, 7500 AE Enschede, The Netherlands}
\homepage{www. photonicbandgaps. com}

\date{Prepared on Sept 9th, 2014}

\begin{abstract}
We study a hybrid system consisting of a narrowband atomic optical resonance and the long-range periodic order of an opaline photonic nanostructure. 
To this end, we have infiltrated atomic cesium vapor in a thin silica opal photonic crystal. 
With increasing temperature, the frequencies of the opal's reflectivity peaks shift down by $> 20\%$ due to chemical reduction of the silica. 
Simultaneously, the photonic bands and gaps shift relative to the fixed near-infrared cesium D$_1$ transitions. 
As a result the narrow atomic resonances with high finesse ($\omega/\Delta \omega = 8 \cdot 10^5$) dramatically change shape from a usual dispersive shape at the blue edge of a stop gap, to an inverted dispersion lineshape at the red edge of a stop gap. 
The lineshape, amplitude, and off-resonance reflectivity are well modeled with a transfer-matrix model that includes the dispersion and absorption of Cs hyperfine transitions and the chemically-reduced opal. 
An ensemble of atoms in a photonic crystal is an intriguing hybrid system that features narrow defect-like resonances with a strong dispersion, with potential applications in slow light, sensing and optical memory.
\end{abstract}

\maketitle
\vspace{1cm}

%%%%%%%%%%%%%%%%%%%%%% Introduction %%%%%%%%%%%%%%%%%%%%%%%%%%%%%

\section{Introduction}
There is great and ongoing interest in nanophotonics to create arrays of resonant systems with high quality factors $Q$ in order to control the propagation and emission of light at a deep and fundamental level. 
New states occur when the state from an individual resonator interacts with the others in the array, such as hybrid collective states or states typical for tight-binding systems~\cite{Ashcroft1976}. 
It has been predicted that periodic arrays of coupled cavities will form waveguides with unusual and tunable dispersion properties~\cite{Yariv1999}. 
If a composite quantum resonance is considered that consists of a cavity strongly coupled to a single quantum dot, such a system will exhibit extreme non-linear behavior in response to even a single energy quantum; arrays of such resonating systems have been predicted to exhibit unusual repulsive boson behavior for individual photons~\cite{Hartmann2006, Greentree2006}. 
Experimentally, great advances have recently been made in the design and fabrication of high-$Q$ cavities in photonic crystals~\cite{Akahane2003, Kuramochi2006, Dharanipathy2013} that may be assembled into arrays of resonators~\cite{Altug2004, Notomi2008}. 
It is very difficult, however, to achieve resonant transport in such an array, since all resonators must mutually be tuned to within their linewidth. 
This entails a tuning to typically one part in a million, which is extremely challenging with the state-of-the-art in nanotechnology. 
Therefore, we propose a new hybrid system exploiting alkali atoms in the gas phase as high-Q resonators in the photonic crystal environment.

Atomic resonances hold several advantages, firstly they exhibit strong and extremely narrow resonances with relative inverse linewidths -  or effective quality factors $Q$ - in the range of $10^7$ and even higher, and secondly their physical properties are extremely well understood. 
Thirdly and most importantly, all atoms (of the same isotope) are identical, hence they do not need to be individually tuned. 
This advantage is in contrast to engineered resonant nanostructures - such as cavity arrays - where each individual unit differs from all other ones~\cite{Koenderink2005}. 

%%%%%%%%%%%%%%%%%% figure 1 %%%%%%%%%%%%%%%%%%%%%%%%%%%%%%%%%%%
\begin{figure}[hbt]
\includegraphics[width=1.0\columnwidth]{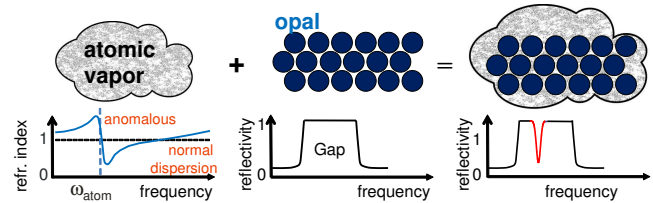}\\
\caption{(color online) Illustration of the concept of our study: 
We study an atomic vapor with a sharp (dipole-allowed) resonance at frequency $\omega_{\rm atom}$ and concomitant dispersion. 
The vapor is introduced in a photonic crystal with a photonic gap such as an opal. 
As a result, we anticipate narrowband resonances inside the photonic gap, shown as the red trough. 
}
\label{fig:concept}
\end{figure}
%%%%%%%%%%%%%%%%%% figure 1 %%%%%%%%%%%%%%%%%%%%%%%%%%%%%%%%%%%
%
To have resonating atoms interact with the optical properties of the host photonic crystal, an ensemble of many atoms must be infiltrated in the crystal, see figure~\ref{fig:concept}. 
The relative dielectric function $\epsilon(\omega) = \epsilon'(\omega) + i \epsilon''(\omega)$ of the atomic medium changes drastically around a resonance frequency $\omega_{a}$, $\epsilon'(\omega)$ exceeding unity below resonance, and being below unity above the resonant frequency $\omega_a$. 
To achieve $\epsilon'(\omega) < 1$, the atomic transitions must be isolated. 
For this reason, we have chosen to study alkali atoms. 
It is desirable for the atomic medium to be dense, in order to have $\epsilon'(\omega)$ change considerably near the resonance frequency $\omega_{\rm atom}$. 
Therefore, we have chosen cesium, as it has the highest vapor density. 
Since the imaginary part of $\epsilon(\omega)$ - associated with extinction - is related to the dispersion by the Kramers-Kronig relations~\cite{Bohren1983}, $\epsilon'(\omega)$ can considerably differ from unity, while $\epsilon''(\omega) << 1$ for a select frequency range outside resonance, see Fig.~\ref{fig:concept}. 
Thus, atomic systems have a large modulation of $\epsilon(\omega)$, which warrants a study of their effect on photonic materials.

It has already been realized that hybrid atom-photonic systems offer exciting prospects: The creation of new modes in the photonic band gap has been predicted~\cite{Coevorden1996, Sivachenko2001}, as well as induced transparency~\cite{Ye2008, Tidstrom2010}, nonlinearities~\cite{Soljacic2004}, and modified pulse propagation~\cite{Camacho2007}. 
Previously, only few experiments have been reported of composite systems of photonic crystals with embedded resonant media. 
It was observed that a strongly absorbing dye in a photonic crystal leads to intriguing braggoriton resonances~\cite{Eradat2002}. 
Unfortunately, however, the linewidth of dye is limited by quenching interactions with other dye molecules that increase with density~\cite{Imhof1999}, thus strongly limiting resonant effects. 
While semiconductor quantum dots are being pursued as internal probes inside photonic crystals, their broad inhomogeneous width (e.g., $5 \%$ for CdSe~\cite{Norris1996}) precludes large changes in $\epsilon'(\omega)$. 
To reduce inhomogeneous width effects, van Coevorden {\it et al.} theoretically studied a photonic crystal consisting of cold atoms on a lattice~\cite{Coevorden1996}. 
Such systems have indeed been studied, although the low lattice site occupancy of $<1 \%$ limited the atomic density in the pioneering studies~\cite{Weidemuller1995, Birkl1995}. 
A 3D band gap is predicted to open when the resonance wavelength equals the lattice spacing~\cite{Coevorden1996}.
The required condition of one atom per lattice site has been realized~\cite{Greiner2002}, although photonic crystal properties have not been reported to date. 
Recently, sub-Doppler features were reported in the spectra of Cs infiltrated in opal nanostructures~\cite{Ballin2013}, and a preliminary study of photonic crystals infiltrated with resonant alkali atoms has been reported~\cite{Harding2008}. 
Here we report on an experiment in which we infiltrate an alkali gas - a saturated cesium vapor - in a thin opal photonic crystal. 

We have decided to study a dense and hot vapor of $^{133}$Cs as a strongly polarizable resonant medium, as it has only one electron in the outer shell in the $6^2S_{1/2}$ state. 
The well-known D-transition from the ground state to the $6P$ manifold is strong, in particular the D$_1$ transitions of $^{133}$Cs with a frequency of 335.116 THz (wavelength $\lambda = 895$ nm) that consist of four isolated hyperfine transitions. 
The excited state has a narrow intrinsic linewidth with $\Gamma_0 = 2 \pi \times 4.56$ MHz, and a large polarizability at resonance~\cite{Young1994, Rafac1998, Rafac1999}. 
In the ground state, the well-known hyperfine splitting of 9.193 GHz defines the unit of time.

\section{Experimental section}
%%%%%%%%%%%%%%%%%%%%%% Setup  %%%%%%%%%%%%%%%%%%%%%%%%%%%%%%

%%%%%%%%%%%%%%%%%% figure 2 %%%%%%%%%%%%%%%%%%%%%%%%%%%%%%%%%%%
\begin{figure}[hbt]
\includegraphics[width=1.0\columnwidth]{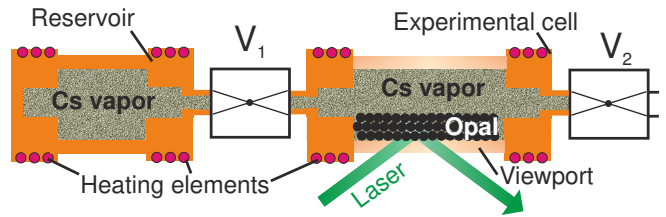}\\
\caption{(color online) 
The heart of the experimental setup is the cesium cell which consists of a Cs reservoir and the measuring cell which contains an opal grown on a vacuum viewport.
These two cells are situated in a large vacuum chamber (not shown) to prevent convection. 
Valves V$_1$ and V$_2$ connect the reservoir with the cell and the cell with the pump, respectively. }
\label{fig:setup}
\end{figure}
%%%%%%%%%%%%%%%%%% figure 2 %%%%%%%%%%%%%%%%%%%%%%%%%%%%%%%%%%%
%
The need for a high vapor density requires heating of liquid Cs. 
Therefore, a cell was built in which Cs can both be heated as well as spatially confined, see Fig.~\ref{fig:setup}. 
Heating is accomplished by two heating elements (thermocoax) brazed at the top and the bottom into the cylindrical reservoir.
Upon the opening of valve V$_1$ (Swagelok SS-4BG-V51, heated), hot Cs vapor diffuses into the experimental cell, which is also heated by two heating elements. 
Because Cs reacts strongly with both oxygen and water, the reservoir and experimental cell are kept in a vacuum of $<10\,\mu$bar, that is detection limited by the low conductance (0.08\,l/s) of the valve and the thin tube. 
To remove moisture and oxygen, the assembly is baked and evacuated at $150\,\degree$C via valve V$_2$ by a turbo-molecular pump for at least 24 hours before admitting Cs. 
Both the reservoir and the experimental cell are independently temperature controlled. 
The temperature is measured by ring J-type thermocouples. 
Heating tape is wound around the tubes and valves to prevent cold spots that would otherwise determine the vapor pressure of Cs. 
The experimental cell, the valves, and the tubes are kept at a temperature 30 to 50\,K above that of the reservoir to prevent condensation. 
The experimental cell is placed in a vacuum tank kept at a pressure below $10^{-3}\,$mbar to prevent convection and thus stabilize the temperature.

To avoid light absorption from Cs vapor between the photonic crystal and the window, \SiOtwo\ opals were grown directly on the viewport windows by vertical controlled drying~\cite{Jiang1999, Hartsuiker2008}. 
To this end, the viewports (VacGen, UK) are cleaned overnight in a solution of 30\,g NaOH, 30\,ml of ultrapure water, and 200\,ml of ethanol. 
All glassware is rinsed in deionized water and in ethanol, and dried in a stream of nitrogen. 
An ethanol suspension of 0.1 to 0.5 vol\,\% \SiOtwo\ colloidal nanospheres with a diameter $D = 460$ nm is placed in a vial. 
The viewport is inserted at an angle between $40\degree$ and $60\degree$ into the vial, which is heated in an oven between $30\degree$C and $60\degree$C. 
Upon evaporation of the suspension, the colloids move towards the area of fastest evaporation, which is the liquid meniscus, and are deposited on the substrate. 
The structure of such an array is close packed, typically face-centered cubic (fcc)~\cite{Miguez1997, Megens2001}. 
As the evaporation proceeds, ordered arrays of spheres grow until finally there is no suspension left and the growth stops~\cite{Jiang1999, Hartsuiker2008}. 
Despite consistently scanning the parameter space in temperature, angle, and colloid density, the opals on the viewports were of moderate quality, with a reflectivity of $2 \Rglass \simeq 10 \%$, where $\Rglass$ is the reflectivity of viewport window. 
In contrast, similar opals grown on microscope slides showed peaks in excess of $13 \Rglass \simeq 55\%$. 
Thus the quality of the opals was limited by the surface quality of the viewports. 
From optical microscope images, the structures were estimated to be between 1 and 3 lattice spacings (i.e., fcc 111 spacings) thick. 
Since we observed greenish opalescence from the deposited nanostructures, likely related to second-order Bragg diffraction (see below), we refer to such a structure as opal. 

The light for the high-resolution measurements is focused at an angle of $\theta = (27\pm2)\degree$ onto the experimental cell, as shown in Fig.~\ref{fig:setup}. 
At this angle the stop bands of the opal are blue shifted to match with the Cs resonance. 
Moreover, this procedure avoids the 2\,mm thick window to act as a Fabry-P\'erot etalon, with a fringe spacing of 
$1/(2\times 0.2\,{\rm cm}\ n_{\rm glass}) = 1.7\,$cm$^{-1}$, or 50\,GHz, which is comparable to the scan range of the laser. 

%%%%%%%%%%%%%%%%%% figure 3 %%%%%%%%%%%%%%%%%%%%%%%%%%%%%%%%%%%
\begin{figure}[hbt]
\includegraphics[width=1.0\columnwidth]{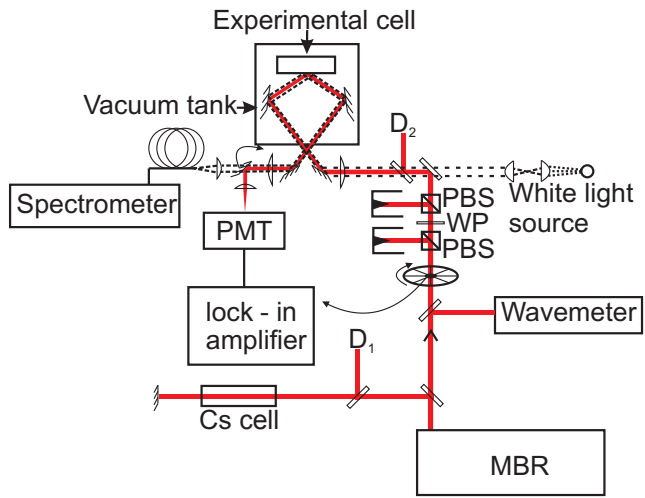}\\
\caption{(color online) 
Optical setup with the optical paths indicated in red.
The narrowband tunable laser (MBR) emits linearly polarized light that is modulated by a chopper.
The laser power is set by polarizing beamsplitters (PBS) and a $\lambda/2$ waveplate (WP), and is monitored by diode D$_2$.
Light reflected by the sample in the experimental cell is detected by a photomultiplier tube (PMT) and lock-in amplifier. 
To determine the optical frequency of the laser, a coarse measurement is performed with a wavemeter, and a high-resolution frequency is measured by saturated absorption spectroscopy in a Cs cell with diode D$_1$. 
Light from a white halogen lamp is used to measure the broadband optical properties of the sample. 
The sample cell is placed in an evacuated tank (volume $0.037\,$m$^{3}$) to minimize air convection around all heated elements. }
\label{fig:setup_large}
\end{figure}
%%%%%%%%%%%%%%%%%% figure 3 %%%%%%%%%%%%%%%%%%%%%%%%%%%%%%%%%%%

Light for the high-resolution experiments is generated by a narrowband single-mode cw laser (Coherent MBR-110), pumped by a frequency doubled Nd:YAG laser (Coherent Verdi V-10), see Fig.~\ref{fig:setup_large}. 
The laser can be scanned over 40\,GHz near $\lambda = 895$ nm, sufficient to excite all four Cs hyperfine transitions in one scan. 
The laser light is linearly polarized, modulated by a chopper, and the power is set by polarization means. 
The beam is focused by an achromatic lens ($f=300\,$mm) at an angle of $\theta = (27\pm2)\degree$ onto the experimental cell containing the cesium vapor and the opal. 
The elastically reflected light is collected by a photomultiplier tube (Hamamatsu R928). 
The photomultiplier is essentially insensitive to fluorescence from the sample due to its very small solid angle of $\Omega\approx 10^{-6}\,$sr. 
The photocurrent of the photomultiplier is measured by phase-sensitive detection.
The laser frequency is coarsely monitored by a wavemeter (Burleigh WA-10L) with a resolution of 0.1\,cm$^{-1}$. 
The precise frequency is measured by Doppler-free spectroscopy~\cite{Demtroder} with a resolution of 26 MHz using a Cs reference cell~\cite{Harding2008}. 
Broadband photonic crystal properties are probed with a spatially-filtered halogen white light source and the reflectivity is measured with a 2048-channel spectrometer (Ocean Optics, USB 2000) in the visible and near infrared range.

\section{Results}
\subsection{Opal features}
%%%%%%%%%%%%%%%%%% figure 4 %%%%%%%%%%%%%%%%%%%%%%%%%%%%%%%%%%%
\begin{figure}[hbt]
\includegraphics[width=\columnwidth]{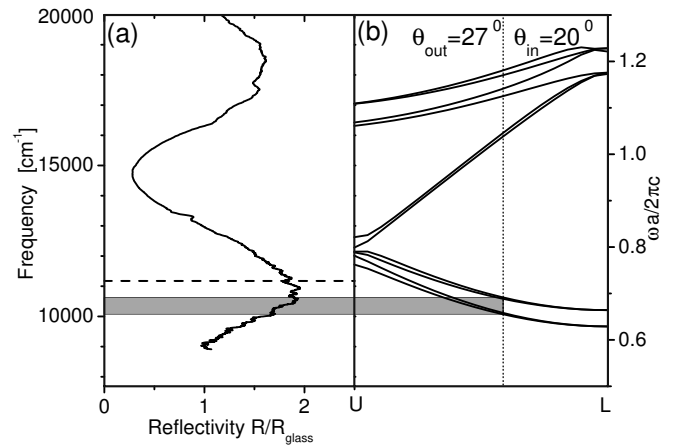}\\
\caption{(a) Reflectivity spectrum of an SiO$_2$ opal grown on the viewport normalized to the reflectance $R_{\rm glass}$ of the viewport, at oblique incidence ($27 \pm 2 ^{\circ}$, internally $20 \pm 2 ^{\circ}$). 
The first peak at $10800$ cm$^{-1}$ is identified with the first-order (111) stop gap of a SiO$_2$ opal. 
The second reflectivity peak is attributed to the range of flat bands and higher-order stop gaps. 
The horizontal dashed line indicates the resonance frequency of the Cs D$_1$ transition. 
(b) Bandstructure from U to L in the irreducible part of the Brillouin zone of an fcc opal photonic crystal. 
The gray horizontal bar indicates the forbidden gap at an angle of $20 \pm 2^\circ$ inside the crystal. 
}
\label{fig:Reflectivity_and_bandstructure}
\end{figure}
%%%%%%%%%%%%%%%%%% figure 4 %%%%%%%%%%%%%%%%%%%%%%%%%%%%%%%%%%%
%
Figure~\ref{fig:Reflectivity_and_bandstructure}(a) shows the broadband reflectivity spectrum of a thin opal grown on a viewport. 
A broad peak is observed at $10800$ cm$^{-1}$ (324\,THz). 
The large relative bandwidth of 48 $\%$ is attributed to finite-size effects of the thin opal~\cite{Bertone1999}, also in view of the low reflectivity. 
At $11178$ cm$^{-1}$, the Cs D$_1$ resonance is indicated, which is slightly blue shifted compared to the first stop band, as aimed for. 
A second high-frequency peak is observed near $18500$ cm$^{-1}$ or 555\,THz.
The high-frequency peaks seems to be split into 3 separate peaks that likely originate from multiple stop-gap interferences and concomitant flat bands, observed earlier in opals and inverse opals~\cite{Vos2000, Romanov2001, Galisteo-Lopez2004}. 
The bandstructure from U to L for a SiO$_2$ opal (n' = 1.45) is shown in Fig.~\ref{fig:Reflectivity_and_bandstructure}(b). 
The peak at $10800$ cm$^{-1}$ corresponds well to the frequency range of the directional forbidden stop gap at an angle of $20 \pm 2^\circ$ inside the crystal. 
The broad high-frequency peak at $18500$ cm$^{-1}$ corresponds well with the range of second-order Bragg diffraction that includes complex stop gaps and flat bands to which external light cannot couple.

%%%%%%%%%%%%%%%%%% figure 5 %%%%%%%%%%%%%%%%%%%%%%%%%%%%%%%%%%%
\begin{figure}[hbt]
\includegraphics[width=\columnwidth]{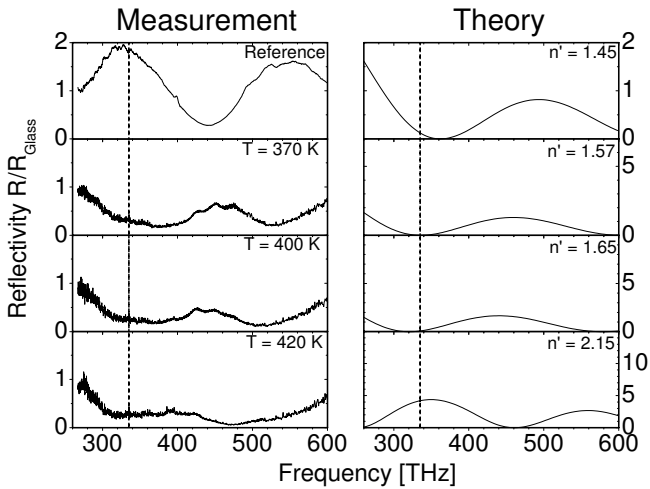}\\
\caption{The stopbands of the opal shift with the increasing temperature because of the progressing reduction of the \SiOtwo. 
On the left the reflection of the opal is shown as a function of probe frequency for four different temperatures, where the first line of spectra is a reference at room temperature without Cs. 
The dashed arrow indicates the red-shifting second-order stop band. On the right, the corresponding transfer-matrix calculations are shown with the used index of the opal material. 
The inset is an extreme zoom in into the region where Cs resonances are expected. The temperature was increased by 10 to 20$\,\degree$C per hour. }
\label{StopbandsFig}
\end{figure}
%%%%%%%%%%%%%%%%%% figure 5 %%%%%%%%%%%%%%%%%%%%%%%%%%%%%%%%%%%
%
Broadband reflectivity spectra are shown in Fig.~\ref{StopbandsFig} for increasing temperature and thus cesium vapor pressure.
At first contact with Cs (not shown), even before spectra are observed, the first-order opal peak has shifted down from 324\,THz to 299\,THz (9.5\,\% relative shift), and the second peak has shifted from 555\,THz to 489\,THz (12\,\%). 
At a cell temperature of $T = 370$ K, the first-order peak has already shifted out of the instrumental spectral range, and the second peak has shifted down to 460\,THz. 
At a temperature of $T = 400$ K, the second reflectivity peak has further shifted to 447\,THz, or 20\,\% down compared to the starting state. 
At higher temperatures the peaks cannot be identified unambiguously. 
The overall reflectivity decreases with increasing temperature, which is caused by a thin oil film that settled on the optics in the vacuum tank. 
The shifting of the reflectivity peaks is attributed to the chemical reduction of the \SiOtwo\ nanospheres~\cite{Jahier2001}; since Cs is strongly caustic, it reduces \SiOtwo\ to \SiOx\ \cite{Pascal1958}.
If $x$ varies from $2$ to $0$, the refractive index of the nanospheres will increase from $n^\prime = 1.45$ for \SiOtwo\ to at most $n^\prime = 3.5$ in case of pure Si~\cite{Andalkar2002}. 
If a homogeneous reduction to SiO occurs, the refractive index of the nanospheres will be equal to that of SiO ($n_{\rm SiO} = 2.0\pm0.1$). 
We surmise that the colloidal nanospheres consist of a mixture of \SiOx\ with different $x$, possibly mixed with Cs oxides.
By substituting the increased average refractive index of the opal layer into Bragg's law~\cite{Vos1996}, the peak shifts to a lower frequency.

%%%%%%%%%%%%%%%%%% figure 6 %%%%%%%%%%%%%%%%%%%%%%%%%%%%%%%%%%%
\begin{figure*}[hbt]
\includegraphics[width=2.0\columnwidth]{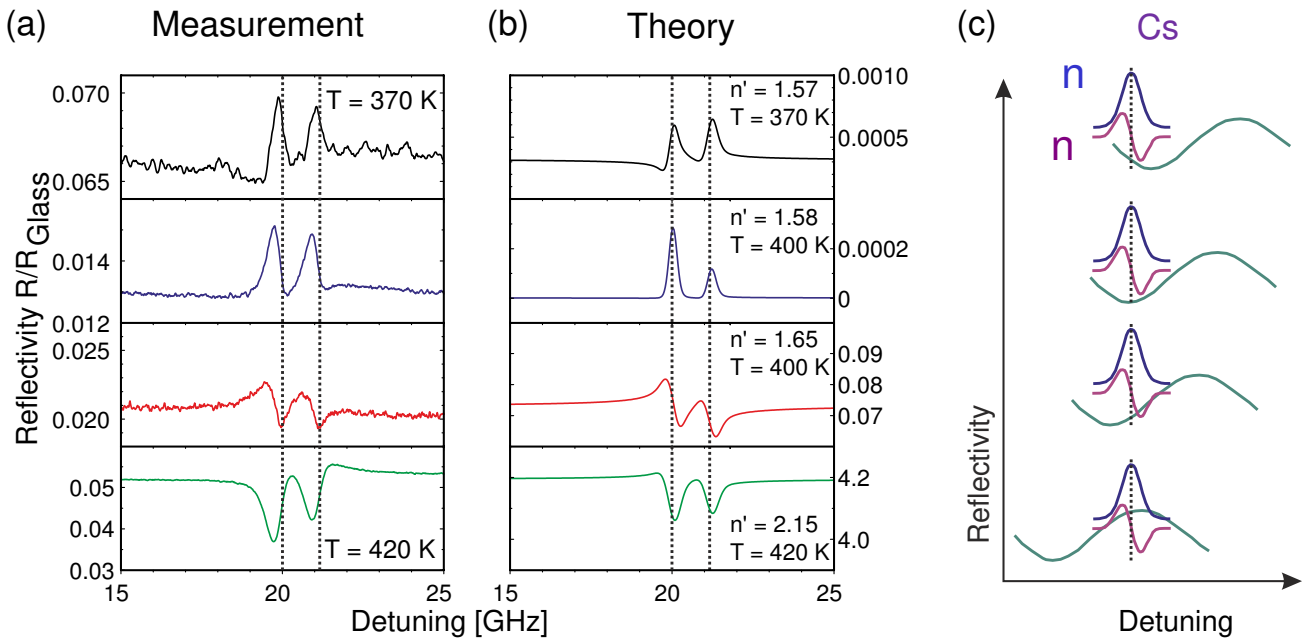}
\caption{(color online) 
(a) High-resolution reflectance data in the frequency range of the Cs D$_1$ transition for consecutive moments in time while the temperature of the cell was slowly increased from 370 K to 420 K (top to bottom), hence increasing Cs densities. 
The frequency of the unperturbed Cs resonances $F=4\rightarrow F^\prime=3$ and $F=4\rightarrow F^\prime=4$ are marked by the dashed lines. 
At a high Cs density (at $420$ K), the Cs resonances appear as red-shifted troughs in reflectivity. 
(b) Calculations with a transfer-matrix model agree well with the reversal of the spectral properties, with indicated nanoparticle refractive indices $n' = 1.57, 1.578, 1.65$, and 2.15 (top to bottom). 
(c) Our result can be understood qualitatively by considering the real and imaginary parts of the reflective index (blue and purple, respectively) and their frequency relative to that of the stopband (green). }
\label{fig:ExplanationFig}
\end{figure*}
%%%%%%%%%%%%%%%%%% figure 6 %%%%%%%%%%%%%%%%%%%%%%%%%%%%%%%%%%%
%
To get more insight in the shape and the shifting of the observed resonances, a transfer-matrix calculation was performed on a single crystal layer of nanospheres surrounded by an atomic vapor, and probed at an angle of $27\degree$. 
At the right hand side of Fig.~\ref{StopbandsFig}, a corresponding transfer-matrix calculation of a single crystal layer is shown. 
The nanosphere's refractive index is chosen to give the best overal match. 
For the reference spectrum, before contact with Cs, the agreement is reasonable (first row). 
The D$_1$ resonance is just blue of the first peak. 
On first contact with Cs, both the first and second peak red shift (not shown). 
At $T = 370\,$K, the second peak has shifted to 460\,THz, which is matched by the calculation for a nanosphere refractive index $n = 1.57$. 
The D$_1$ resonance is at the blue foot of the first-order peak. At $T = 400$K, the D$_1$ resonance is exactly in the trough between two peaks, obtained for $n = 1.65$. 
As the temperature is increased even further, $n = 2.15$ is fitted, shifting the second peak so that the D$_1$ resonance is now on its red side. 

\subsection{Atomic features}
We see that the increase in cesium density (moderated by the increasing temperature) and progression in time increase the nanosphere refractive index. 
As a result, the opal reflectivity peaks are tuned with respect to the atomic resonances. 
Compared to the broadband reflectivity, we increased the spectral resolution by $10^4$ to resolve the Cs hyperfine structure. 
Therefore, intermixed with the measurements of the broadband spectra we have scanned the narrowband excitation laser and recorded the reflected intensity, as displayed in Fig.~\ref{fig:ExplanationFig}, that zoom in on the Cs D$_1$ transitions. 
At a low Cs density (at $370$ K), the background reflection of the opal is low and the Cs resonances appear as red-shifted peaks in reflectivity. 
To determine the linewidth, we modeled the peaks with a Gaussian function and obtain about $\Gamma = 2\pi\times 0.4$ GHz FWHM.
From earlier work~\cite{Harding2008}, we deduced that the peaks are both collision and Doppler broadened. 
If we consider the atoms to behave as tiny resonators, the width correspond to an effective quality factor $Q=\omega/\Delta\omega=8\times10^{5}$. 
While similar - or even higher - quality factors have been reported~\cite{Akahane2003, Kuramochi2006, Dharanipathy2013}, the advantageous feature of the Cs atoms studied here is that their center frequency varies by less than 1 part in $840.000$ (= 335 THz/0.4 GHz), which is beyond state-of-the-art in nanocavity fabrication. 
Therefore, atoms warrant attention as resonant systems in nanophotonic arrays. 

Figure~\ref{fig:ExplanationFig}(a) shows measured high-resolution reflectivity spectra at four moments in time while the temperature and hence the Cs density was raised. 
The two hyperfine resonances $F=4\rightarrow F^\prime=3$ and $F=4\rightarrow F^\prime=4$ are clearly observed. 
At $T = 370$ K, the background reflectivity of the opal is around $0.065 R_{\rm{glass}}$, or $0.26\%$. 
Close to the two resonances, the reflectivity shows two clear peaks. 
Below the $4 \rightarrow 3'$ transition frequency, the spectrum shows a short tail with a trough, and beyond the resonances the spectrum shows a broad and asymmetric tail. 
When the second spectrum is taken, both peaks have become more pronounced.
The low-frequency tail has expanded, and the low-frequency trough has disappeared. 
The background reflectivity of the opal has decreased to 0.013. 
The third spectrum has dramatically changed: after the first peak a trough has appeared that clearly dives below the opal's background. 
Similarly, a second trough with a minimum below the background has appeared at the second resonance. 
The initial trough below resonance has completely disappeared, and the baseline has recovered to 0.02. 
The last spectrum, at $T = 420$ K, shows another striking change: the peaks seen at the earlier spectra have completely vanished and given way to two marked troughs. 
The spectra is completely without peaks, exactly the opposite of the second spectrum. 
Moreover, the opal background has recovered even further to 0.05. 

Our observations can be qualitatively understood by considering a polarizable atomic medium inside a photonic structure, whose reflectivity peaks shift due to a changing opal refractive index. 
These features have been incorporated in our transfer-matrix model, whose results are shown in Fig.~\ref{fig:ExplanationFig}(b). 
We consider the real and imaginary parts of the atomic refractive index $n^\prime$ and $n^{\prime\prime}$, respectively, and their frequency relative to that of the opal's stopband (see Fig.~\ref{fig:ExplanationFig}).
We distinguish 4 characteristic cases as illustrated in Fig.~\ref{fig:ExplanationFig}(c).

\begin{enumerate}
\item{If the atomic resonance is at the blue edge of the reflectivity peak, an increase in $n^\prime$ from unity below resonance will effectivly cause the stop gap to red shift at that frequency. 
A red shift at the blue edge of a peak reduces the reflectivity. 
Above resonance, the opposite happens since $n^\prime$ is below unity. 
The behavior of $n^\prime$ versus frequency across the resonance thus results in an decreasing and increased reflectivity, as seen at $T = 370$ K. 
On resonance, the enhanced $n^{\prime\prime}$ will destroy the destructive interference, thereby increasing the reflectivity. 
}
\item{For resonances at the trough between two opal reflectivity peaks, the increase in $n^{\prime\prime}$ on resonance will dominate the reflectivity and cause the two peaks, as seen on the second row where we fitted n'=1.58. 
The gradient of the opal's reflectivity vanishes, and any change in $n^\prime$ will hardly change $R(\omega)$, since the opal's reflectivity peaks will be little shifted. 
}
\item{At the red edge of the second opal peak, the increase in $n^\prime$ below resonance will red shift the opal peak.
As a result the reflectivity will increase. 
Above resonance, the decrease in $n^\prime$ will blue shift the opal peak, resulting in a decrease of the reflectivity. 
Hence, the measured reflectivity has a positive correlation with $n^\prime$. 
The result of increase in $n^{\prime\prime}$ depends on whether the resonance is at the foot of the opal reflectivity peak or near the maximum: in the former case, the reflectivity increases, while in the latter case the constructive Bragg interference is suppressed, resulting in a reflectivity trough.
} 
\item{An atomic resonance near the center of the reflectivity peak will lead to following behavior: 
Since the gradient of the reflectivity peak is small at its maximum, the induced shift by the change in $n^\prime$ will have little effect on the opal's reflectivity. 
In contrast, the absorption expressed by the $n^{\prime\prime}$ will remove the constructive Bragg interference. 
Hence an increase in $n^{\prime\prime}$ will correlate with a trough at resonance, as seen in the last spectrum, at $T = 420$ K.
}
\end{enumerate}

It appears that the resonance frequencies in our experiments are systematically lower from calculated ones. 
The difference is attributed to a pressure shift of remnant N$_2$\ in the reservoir. 
Using the shift due to a \Ntwo\ buffer gas of -10.97\,MHz/mbar from Ref.~\cite{Andalkar2002}, a partial pressure of 27\,mbar explains the difference. 
This partial pressure agrees excellently with a measurement of $27\pm10\,$mbar at evacuation after the experiment. 

From the good match between the measured spectra and the theory curves we conclude that the system is well described by an atomic vapor filling the voids around the opal spheres. 
Moreover, from the agreement we deduce that the real part of the refractive index must have been less than unity for probe frequencies above resonance. 
To the best of our knowledge, this is the first observation of sub-unity refractive index in a photonic crystal, inside a stop band.

\section{Discussion}
The striking spectral features reported above reveal that a hybridization occurs between the optical properties of the cesium atoms and the opal structure. 
The result of the hybridization depends strongly on the tuning of the atomic resonances relative to the opal's gap structure. 
The simple case anticipated in Fig.~\ref{fig:concept} occurs when the atomic resonances are centred in the gap. 
This case corresponds to the observations at $T = 420$ K in Fig.~\ref{fig:ExplanationFig}. 
In addition, intriguing hybridizations are observed when the atomic resonances are tuned to either the red or the blue edge of a stop gap. 
In these cases, the real parts of the atomic dispersion come into play, leading to asymmetric reflectivity features in the observed spectra. 
This situation differs from arrays of cavities in photonic crystals. 
In the latter case, the cavities must necessarily be tuned deep within the gap, in order to profit from the confinement offered by the gaps.
Conversely, a cavity resonance tuned to the edge of a gap will obviously be lossy, since the cavity resonance is derived from the surrounding photonic crystal, in contrast to an atomic resonance, that obviously also exists outside a photonic crystal. 

In contrast to the one-dimensional (1D) cases of long hollow optical fibers~\cite{Muller2000, Ghosh2006,Light2007,Slepkov2008} or waveguides on a chip~\cite{Yang2007}, three-dimensional (3D) photonic crystals have the demonstrated potential to significantly alter the local density of states (LDOS)~\cite{Lodahl2004, Noda2007, Leistikow2011}, allowing to control the excited-state lifetime of emitters more in 3D~\cite{Vats2002} than in 1D~\cite{Hooijer1999}.
Our results demonstrate that the required high atomic densities can be achieved. 
Several improvements merit investigation in future studies. 
First of all, other combinations of gases and photonic crystals will have to be more stable against chemical interactions.
Another option is to coat the opal with a thin protective layer or to use light-induced adsorption to increase the alkali pressure. 
Once these issues are resolved, we can investigate the influence of the LDOS on the lifetime of excited states.
The high-$Q$ resonances in a photonic crystal could even allow to study the thermalization of a photon gas~\cite{Klaers2010}. 
Finally, our experiment could offer an alternative experimental route to complex experiments that are being done with atomic lattices~\cite{Bloch2008}.

The measured peaks are somewhat broader than those calculated with our model. 
The model does not take into account collisional broadening or transit time broadening. 
While collisional self-broadening~\cite{Chen1968} does not play a significant role at our experimental conditions, the N$_2$\ background broadens a delta-function line to about 400\,MHz FWHM~\cite{Andalkar2002}. 
%The Cs atoms fly in very narrow holes between the solid \SiOx\ spheres of 230\,nm radius. Taking this radius as characteristic flight length, and taking the average thermal velocity in a certain direction (i.e., $\alpha/\sqrt{\pi}$, where $\alpha$ the most probable speed in the gas) as characteristic velocity this should lead to an additional broadening of 0.42 to 0.53\,GHz for temperatures of 300 to 473\,K, respectively~\cite{Demtroder}. 
In addition, the atoms between the opal spheres will have only a small interaction time with the light before hitting a material wall or before being outside the optical focus, leading to transit-time broadening~\cite{Demtroder}. 
Since the magnitude strongly depends on geometry, it is hard to estimate, yet it could easily amount to a few 100\,MHz. 
In view of these broadening mechanisms, the observed resonances are surprisingly narrow, suggestive of selective reflection effects~\cite{SelectivereflectionLit, Wang1997}. 
In selective reflection, slow atoms with only a small velocity component parallel to the propagation direction of light contribute disproportionately to the signal, leading to sharp spectral features. 
The prospect of exploiting selective reflection to yield even sharper features in future makes this system even more attractive. 
 
\section{Conclusions}
We report on an experiment in which we infiltrate a dense alkali gas - a saturated cesium vapor - in a thin opal photonic crystal. 
The stop bands of the nanostructure are tuned to overlap with the narrow atomic resonances of the alkali atoms, leading to an intriguing hybridization of the atomic resonances and the crystal spectrum. 
As the temperature is increased, we observe a shift of the peak opal reflectivity frequency in excess of $20\,\%$, which is attributed to the chemical reduction of the silica by the cesium. 
As a result, the frequency of the photonic bands are scanned through the near-infrared Cs D$_1$-transition. 
Simultaneously, the Cs resonances undergo marked changes in strength, off-resonance reflectivity, and lineshape. 
When the atomic resonance is tuned to the red edge of the stop gap, the reflectivity reveals a striking inversely dispersive shape. 
We successfully interpret all observed spectral features by using a transfer-matrix model that includes the dispersion and absorption of two of the hyperfine transitions of Cs, as well as the ongoing change of the opal’s refractive index due to the chemical reduction. 
Our study is a step forward in the understanding of strongly resonant systems inside photonic crystals, and shows that a hybrid combination of atomic and condensed matter holds interesting physics.

\section{Acknowledgments}
It is a pleasure to thank Cock Harteveld, and Leon Woldering for expert technical support. 
We thank Javier Garc{\'i}a de Abajo, Tom Hijmans, Ad Lagendijk, and Gerhard Rempe for encouragements on the project. 
This work is part of the research program of the 'Stichting voor Fundamenteel Onderzoek der Materie' (FOM) that is financially supported by the 'Nederlandse Organisatie voor Wetenschappelijk Onderzoek' (NWO), notably the PPOM program and “Inrichting leerstoelpositie” grant (02ILP012). 
We also thank ERC (grant no 279248), NWO (Vici), and STW for support.

%%%%%%%%%%%%%%%%%%%%%%%%%%%%%%% References %%%%%%%%%%%%%%%%%%%%%%%%%%%%%

%%%%%%%%%%%%%%%%%%%%%%%%%%%%%%%%%%%%%%%%%%%%%%%%%%%%%%%%%%%

\end{document}